\documentclass[conference,letterpaper]{IEEEtran}

\usepackage[utf8]{inputenc} 
\usepackage[T1]{fontenc}
\usepackage{url}
\usepackage{ifthen}
\usepackage{cite}
\usepackage[cmex10]{amsmath} 

\usepackage{graphicx}
\usepackage{amssymb}

\usepackage{amsmath}
\usepackage{bm}
\usepackage{color}

\newtheorem{theorem}{Theorem}

\newtheorem{lemma}{Lemma}
\newtheorem{definition}{Definition}

\newtheorem{remark}{Remark}
\newtheorem{example}{Example}

\newcommand{\triequ}{\stackrel{{\scriptscriptstyle{\bigtriangleup}}}{=}}
\newcommand{\myQED}{\hfill{$\Box$}}
\newcommand{\fat}[1]{\mbox{\boldmath$#1$}}
\newcommand{\Eqn}[1]{&\hspace{-0.5em}#1\hspace{-0.5em}&}
\newcommand{\redunderline}[1]{{\textcolor{black}{#1}}}
\newcounter{xyz}

\begin{document}

\title{A construction of UD $k$-ary multi-user codes from $(2^m(k-1)+1)$-ary codes for MAAC}

 \author{%
   \IEEEauthorblockN{Shan Lu\IEEEauthorrefmark{1},
                     Wei Hou\IEEEauthorrefmark{2},
                     Jun Cheng\IEEEauthorrefmark{3},
                     and Hiroshi Kamabe\IEEEauthorrefmark{1}}
  \IEEEauthorblockA{\IEEEauthorrefmark{1}%
                     Department of Electrical,
                     Electronic, and Computer
                     Engineering, Gifu University, Gifu, 501-1193 Japan}
   \IEEEauthorblockA{\IEEEauthorrefmark{2}%
                     The State Key Laboratory of ISN, Xidian University, Xi'an, 710071, China}
   \IEEEauthorblockA{\IEEEauthorrefmark{3}%
                     Department of Intelligent Information Engineering and Sciences, Doshisha University, Kyoto, 610-0321 Japan}
 }

\maketitle

\begin{abstract}
In this paper, we proposed a construction of a UD $k$-ary $T$-user coding scheme for MAAC.
We first give a construction of $k$-ary $T^{f+g}$-user UD code from a $k$-ary $T^{f}$-user UD code and a $k^{\pm}$-ary $T^{g}$-user difference set with its two component sets $\mathcal{D}^{+}$ and $\mathcal{D}^{-}$ {\em a priori}. Based on the $k^{\pm}$-ary $T^{g}$-user difference set constructed from a $(2k-1)$-ary UD code, we recursively construct a UD $k$-ary $T$-user codes with code length of $2^m$ from initial multi-user codes of $k$-ary, $2(k-1)+1$-ary, \dots, $(2^m(k-1)+1)$-ary. Introducing multi-user codes with higer-ary makes the total rate of generated code $\mathcal{A}$ higher than that of conventional code.
\end{abstract}

\section{Introduction}
In a multiple access communication system, $T$ independent sources transmit data over a common channel. A central problem in multi-user coding for the multiple-access communications system is to assign $T$ constituent codes $\mathcal{C}_j$ 
($j=1,2,\cdots,T$) to $T$ senders so that they can communicate simultaneously with a common receiver through a multiple-access adder channel(MAAC),
even in the presence of noise. 
The collection of $T$ constituent codes $\mathcal{C}_j$ is the $T$-user $k$-ary code, denoted by $\mathcal{C}=\{ \mathcal{C}_1,  \mathcal{C}_2, \dots, \mathcal{C}_T \}$. 

The $T$-user $k$-ary code is said to be {\em uniquely decodable } (UD) if all sums consisting of one codeword from each constitute code are distinct. Most extensively investigated UD multiuser codes for the MAAC are binary \cite{Lindstrom75} \cite{Martirosyan89} \cite{Chang79} \cite{Hughes96}. Lindstr\"{o}m, Martirosyan, and Khachatryan gave the multiuser UD codes, where each constituent code consists of only two codewords, one of them being zero \cite{Lindstrom75} \cite{Martirosyan89}. Chang and Weldon's construction also focused on the symmetric case of each constituent code with two codewords \cite{Chang79}. 

By distributing those codewords of \cite{Martirosyan89} or \cite{Chang79} among as few users as possible, affine multiuser codes are proposed\cite{Hughes96} \cite{Khachatrian98} to achieve higher total rate for a fixed number of users. Each constituent code Hughes and Cooper's code \cite{Hughes96} consist of $2^m$ codewords and Martirosyan's code \cite{Khachatrian98} is free of that constraints and achieves the higher total rate for a fixed number of users than those of \cite{Martirosyan89} \cite{Chang79}.

Multiuser $k$-ary ($k \geq 3$) UD codes was originally studied by Jevti\'{c} \cite{Jevtic95} and was then generalized to the case of arbitrary code length in \cite{Cheng99}. By extension of Hughes and Cooper's idea \cite{Hughes96} to $k$-ary, a mixed-rate multiuser $k$-ary code was given in \cite{Cheng00_AffineUD} and achieves the higher total rate for a fixed number of users than that of \cite{Jevtic95} \cite{Cheng99}.

On the other hand, error-correcting multi-user codes are also studied for noisy MAAC \cite{Chang79}\cite{Wilson88}\cite{Cheng00}\cite{Cheng01}. They usually construct the error-correcting codes from the UD codes.

\redunderline{
In this paper, we proposed a construction of a UD $k$-ary $T$-user coding scheme for MAAC.
We first give a construction of $k$-ary $T^{f+g}$-user UD code from a $k$-ary $T^{f}$-user UD code and a $k^{\pm}$-ary $T^{g}$-user difference set with its two component sets $\mathcal{D}^{+}$ and $\mathcal{D}^{-}$ {\em a priori}. Based on the $k^{\pm}$-ary $T^{g}$-user difference set constructed from a $(2k-1)$-ary UD code, we recursively construct a UD $k$-ary $T$-user codes with code length of $2^m$ from initial multi-user codes of $k$-ary, $2(k-1)+1$-ary, \dots, $(2^m(k-1)+1)$-ary. Introducing multi-user codes with higer-ary makes the total rate of generated code $\mathcal{A}$ higher than that of conventional code.
}

\section{Preliminary}

\subsection{$T$-user $k$-ary MAAC}
Let ${\cal K}\!\triequ\!\{0,1,2,\cdots,k-1\}$ for a given positive integer $k$ not less than $2$.
Denote $U_j\! \in\! {\cal K} (j=1,2,\cdots,T)$ the $j$th user's input, the output of the {\em noiseless MAAC} at each time epoch is given by $$Y\!=\! \sum_{j=1}^{n}U_j,$$ where summation is over
the real numbers. Clearly, each output symbol of the channel is an integer from set $\{0,1,2,\!\cdots,\!(k-1)T\}$. Symbol and block synchronization are assumed and all of the users are always active.

A {\em noisy MAAC} is regarded as a noiseless MAAC cascaded with a discrete memoryless channel \cite{Chang79}. The discrete memoryless channel is $((k-1)T+1)$-ary-input and $((k-1)T+1)$-ary-output, and is completely described by transition probabilities for all the possible input-output pairs $(i,j)$, $0 \leq i,j \leq (k-1)T$.

\subsection{$T$-user code}
Let $\mathcal{C}=\{\mathcal{C}_1, \mathcal{C}_2, \cdots, \mathcal{C}_T\}$ be a $T$-user $k$-ary code for MAAC composed of constituent codes $\mathcal{C}_j \subset {\cal K}^n$, $j=1,2,\cdots,T$. The {\em rate} of $\mathcal{C}_j$ is $R_j=(1/n)\log_2|\mathcal{C}_j|$,
where $|\mathcal{C}_j|$ is the cardinality of $\mathcal{C}_j$. The {\em total rate} of $\mathcal{C}$
is $R(\mathcal{C})\triequ R_1+R_2+\cdots+R_T$.

The {\em weight} of an integer $n$-vector (a row vector with length
$n$) ${\boldsymbol x}=(x_1,x_2,\cdots,x_n)$ is defined as
${w}({\boldsymbol x})=\sum_{j=1}^{n}|x_j|$, where summation is over
the real numbers. The {distance} between two vectors
${\boldsymbol y}$ and ${\boldsymbol y}^\prime$ is defined as
$d({\boldsymbol y},{\boldsymbol y}^\prime)={w}({\boldsymbol
y}-{\boldsymbol y}^\prime)$, where the minus sign ``$-$'' denotes
componentwise real-number subtraction.

\begin{definition}\cite{Chang79}{\label{def:MinDis}}
For any positive integer $\delta$, a $T$-user $k$-ary code $\mathcal{C}=\{\mathcal{C}_1,\mathcal{C}_2,\cdots,\allowbreak \mathcal{C}_T\}$ over $\mathcal{K}^{n}$ with code
length $n$ is ${\delta}$-{\em decodable}, if for any two distinct
$k$-ary $nT$-vectors $({\boldsymbol u}_1,{\boldsymbol
u}_2,\cdots,{\boldsymbol u}_T)$ and
$({\boldsymbol u}_1^{\prime},{\boldsymbol
u}_2^{\prime},\cdots,{\boldsymbol u}_T^{\prime})$, ${\boldsymbol u}_j, {\boldsymbol
u}_j^{\prime} \in \mathcal{C}_j$,
\begin{eqnarray}
d \left( \sum_{j=1}^{T}{\boldsymbol u}_j, \sum_{j=1}^{T}{\boldsymbol u}_j^{\prime} \right)
\geq \delta.
\end{eqnarray}
\end{definition}

A $T$-user $\delta$-decodable $k$-ary code $\mathcal{C}$ with code length $n$ is referred to as an $(n,\delta,T)_k$-{\em code}.  A ${\delta}$-decodable
code is capable of correcting $\lfloor ({\delta}-1)/2 \rfloor$ or fewer
errors \cite{Chang79}, where $\lfloor p \rfloor$ is the greatest integer
less than or equal to $p$. On the other hand, the $\delta$-decodable
code is used for detecting $\delta-1$ or fewer errors.

An $(n,\delta=1,T)_k$-code is said to be {\em uniquely decodable} (UD), and is used for the noiseless MAAC such that the decoder is able to uniquely resolve any possible received word into the transmitted codewords, one from each of the users.

\begin{lemma}{\label{le:UD}}
A $T$-user $k$-ary code $\mathcal{C}=\{\mathcal{C}_1,\mathcal{C}_2,\cdots,\allowbreak \mathcal{C}_T\}$ is UD, if for any two 
$k$-ary $nT$-vectors 
$({\boldsymbol u}_1,{\boldsymbol u}_2,\cdots,{\boldsymbol u}_T)$ and
$({\boldsymbol u}_1^{\prime},{\boldsymbol u}_2^{\prime},\cdots,{\boldsymbol u}_T^{\prime})$, ${\boldsymbol u}_j, {\boldsymbol u}_j^{\prime} \in \mathcal{C}_j$,
\begin{eqnarray*}
 \sum_{j=1}^{T}({\boldsymbol u}_j- {\boldsymbol u}_j^{\prime})= \fat{0}^n
\end{eqnarray*}
implies 
\begin{eqnarray*}
({\boldsymbol u}_1-{\boldsymbol u}_1^{\prime},{\boldsymbol u}_2-{\boldsymbol u}_2^{\prime},\dots,{\boldsymbol u}_T-{\boldsymbol u}_T^{\prime})=\fat{0}^{nT}.
\end{eqnarray*}
\myQED
\end{lemma}

\subsection{$T_d$-subset difference set}

Let ${\cal K}^{\pm} \triequ \{0,\pm 1, \pm 2,\dots, \pm (k-1)\}$. Denoted by $\mathcal{D}=\{\mathcal{D}_1, \mathcal{D}_2, \dots, \mathcal{D}_{T_d}\}$ an $(n,\delta_d, T_d)_{2k-1}$ code with $\mathcal{D}_j \subset ({\mathcal K}^{\pm})^{n}$. Given vector $\fat{d}_j \in (\mathcal{K}^{\pm})^{n}$, we obtain two vectors $\fat{d}_j^{+}$ and $\fat{d}_j^{-} \in \mathcal{K}^{n}$ in the following manner. Let $d_{jm}\in \mathcal{K^{\pm}}$ be the $m$th element in $\fat{d}_j$, and $d^{+}_{jm}$, $d^{-}_{jm} \in \mathcal{K}$ be the $m$th element in $\fat{d}^{+}_j$, $\fat{d}_j^{-}$, respectively. We can always put
\begin{eqnarray}{\label{d+d-}}
\begin{array}{lll}
d^+_{jm} = d_{jm}, &d^-_{jm} = 0,   & \mbox{if~~}  d_{jm} \geq 0 \\
d^+_{jm} = 0, & d^-_{jm} = |d_{jm}|, & \mbox{if~~}  d_{jm} < 0.
\end{array}
\end{eqnarray}
Let
\begin{eqnarray}{\label{eq:D+D-}} 
\begin{array}{l}
{\mathcal D}^{+}_j =\{\fat{d}^{+}_j |\fat{d}^+_j {\mbox{ subject to (\ref{d+d-})}}, \fat{d}_j \in {\mathcal D}_j \} \\
{\mathcal D}^{-}_j =\{\fat{d}^{-}_j |\fat{d}^-_j {\mbox{ subject to (\ref{d+d-})}}, \fat{d}_j \in {\mathcal D}_j\}.
\end{array}
\end{eqnarray}
We obtain $\mathcal{D}^{+} = \{\mathcal{D}^{+}_1, \cdots, \mathcal{D}^{+}_{T_d}\}$ and $\mathcal{D}^{-} = \{\mathcal{D}^{-}_1, \cdots, \mathcal{D}^{-}_{T_d}\}$. Note that both of sets are over $\mathcal{K}^n$.

Since $\fat{d}_j = \fat{d}_j^{+} - \fat{d}_j^{-}$, for convenience we refer to
the $(n,\delta_d, T_d)_{2k-1}$-code $\mathcal{D}$ as $T_d$-subset $\delta_d$-decodable {\em difference set} over $(\mathcal{K}^{\pm})^{n}$, denoted by $\langle n, \delta_d, T_d \rangle _{\pm k}$. Except for the domain $(\mathcal{K}^{\pm})^{n}$ of the set, the definition of $T_d$-subset difference set is almost same as that in Definition~\ref{def:MinDis}. In this paper, we focus on the code over $\mathcal{K}^{n}$. To avoid confusion, we call $\mathcal{D}$ as a set over $(\mathcal{K}^{\pm})^{n}$, not a code. The difference set over $({\cal K}^{\pm})^n$ plays an important role in our coding scheme.

\section{Uniquely Decodable Multi-User Codes}{\label{eq:kUDCode}}

\subsection{$T^{f+g}$-User UD Code}
First, we give a construction of $T^{f+g}$-user UD code from a $T^{f}$-user UD code and a $T^{g}$-user difference set.

Let $f$ and $g$ be nonnegtive integers. Without loss of generality, we assume $f \geq g$. Let
$$
\mathcal{A}^f
=\{{\mathcal{A}}_1, {\mathcal{A}}_2, \dots, {\mathcal{A}}_{T^f}\},
\quad {\mathcal{A}}_j\ \subset \mathcal{K}^{f}
$$
be $(f,\delta=1,T^f)_k$-code, and
$$
\mathcal{D}^g = \{{\mathcal{D}}_1, {\mathcal{D}}_2, 
\dots, {\mathcal{D}}_{T^g}\},
\quad {\mathcal{D}}_j\ \subset \mathcal{K}^{g}
$$
be $(g,\delta=1,T^g)_{\pm k}$-difference set, respectively. When 
$\mathcal{A}^f$ and $\mathcal{D}^g$
 are given {\em a priori}, a $(T^f+T^g)$-user code 
\begin{eqnarray}{\label{eq:UDfgCode}}
\mathcal{C}^{f+g} 
\triequ {\Omega(\mathcal{A}^f,\mathcal{D}^g)}
=\{{\mathcal{C}}_1, {\mathcal{C}}_2,\dots,{\mathcal{C}}_{T^f+T^g}\}
\end{eqnarray}
is made as follows:
\begin{eqnarray*}
\mathcal{C}_{i} 
\!\!\! \Eqn{=} \!\!\! 
\{\fat{u}_{i}\!\! =(\fat{a},\fat{a}^{(g)})
\ |\ \fat{a}
=(\fat{a}^{(g)},\fat{a}^{\langle f-g \rangle}) 
\in {\mathcal{A}}_i \} \label{eq:UDfCode}\\
&& ~~~~~~~~~~~~~~~~ i=1,2,\dots,T^f \nonumber \\
\mathcal{C}_{i+T^f} 
\!\!\! \Eqn{=} \!\!\! 
\{\fat{u}_{i+T^f}\!\! =(\fat{d}^{+[f]},\fat{d}^{-})\\
&&\vspace{-7mm}|\ \fat{d}^{+[f]} =(\fat{d}^+,\fat{0}^{f-g}),  
\fat{d}^{+} \in {\mathcal{D}}_{i}^+, 
\fat{d}^{-} \in {\mathcal{D}}_{i}^-, \} \label{eq:UDgCode}\\
&& ~~~~~~~~~~~~~~~~ i=1,2,\dots,T^g. \nonumber
\end{eqnarray*}

\noindent The notations in the above equations are defined as follows:
\begin{list}{}{\itemsep=0mm \parsep=0mm \parskip=0mm \topsep=0mm}
\usecounter{xyz}
\renewcommand{\makelabel}{(\alph{xyz})}


\item 
${\fat{a}}^{(g)}$ is the first $g$ components of vector $\fat{a}$, and $\fat{a}^{\langle f-g \rangle}$ is the last $f-g$ components of $\fat{a}$, i.e.,
$\fat{a}=(\fat{a}^{(g)},\ \fat{a}^{\langle f-g \rangle}).$

\item $\fat{d}^{+[f]}$ is an $f$-vector whose first $g$ components are the vector $\fat{d}^+$, and whose last $f-g$ components are ${\boldsymbol 0}^{f-g}$, i.e.,
$
\fat{d}^{+[f]}=(\fat{d}^{+},\ {\boldsymbol 0}^{f-g}).
$
\end{list}

\begin{theorem}{\label{th:UDfgCode}}
If the $k$-ary code $\mathcal{A}^f$ and the difference set $\mathcal{D}^g$ over $({\cal K}^\pm)^n$ are UD, then ${\Omega(\mathcal{A}^f,\mathcal{D}^g)}$ of (\ref{eq:UDfgCode}) is a $(f+g, \delta=1,T^f+T^g)_k$-code.
\myQED
\end{theorem}

{\em Proof}:
Clearly, ${\Omega(\mathcal{A}^f,\mathcal{D}^g)}$ has code length $f+g$, and is $k$-ary. The number of users in ${\Omega(\mathcal{A}^f,\mathcal{D}^g)}$ are $T^f+T^g$.

Let us prove that ${\Omega(\mathcal{A}^f,\mathcal{D}^g)}$ is UD. Let
\begin{eqnarray*}
\fat{u}_i &\in& \mathcal{C}_i,\quad i=1,\dots,T^f \\
\fat{u}_{i+T^f} &\in& \mathcal{C}_{i+T^f}, 
\quad i=1,\dots,T^g
\end{eqnarray*}
By Lemma~\ref{le:UD}, we will show that
\begin{eqnarray}{\label{eq:UDWordDif0}}
\sum_{i=1}^{T^f+T^g}(\fat{u}_{i} -\fat{u}_{i}^{\prime})
=\fat{0}^{f+g}
\end{eqnarray}
implies
\begin{eqnarray} {\label{eq:UDTwoDistinctVectorsEqu}}
&& \hspace*{-4em} 
(\fat{u}_{1}-\fat{u}_{1}^\prime, 
\dots, \fat{u}_{T^f}-\fat{u}_{T^f}^\prime, 
\nonumber\\
&&\quad \fat{u}_{T^f+1}-\fat{u}_{T^f+1}^\prime, 
\dots, \fat{u}_{T^f+T^g}-\fat{u}_{T^f+T^g}^\prime) 
\nonumber\\
&&\qquad = \fat{0}^{(f+g)(T^f+T^g)}.
\end{eqnarray}

We have that
\begin{eqnarray}{\label{eq:UDWordDiff}}
\lefteqn{\sum_{i=1}^{T^f+T^g}(\fat{u}_{i} -\fat{u}_{i}^{\prime})}\\
&=&\sum_{i=1}^{T^f}\left((\fat{a}_{i},\fat{a}_i^{(g)}) 
-(\fat{a}_{i}^\prime,\fat{a}_i^{\prime (g) }) \right) 
\nonumber \\
&& + \sum_{i=T^f+1}^{T^f+T^g}
\left((\fat{d}_{i}^{+[f]},\fat{d}_i^{-}) 
-(\fat{d}_{i}^{\prime +[f]},\fat{d}_i^{\prime-}) \right)
\nonumber\\
&\triequ& (\fat{s}_1,\fat{s}_2).
\end{eqnarray}
with
\begin{eqnarray}
\fat{s}_1 
\Eqn{=} \sum_{i=1}^{T^f}(\fat{a}_{i}-\fat{a}_{i}^\prime)
+\sum_{i=T^f+1}^{T^f+T^g}(\fat{d}_{i}^{+[f]}-\fat{d}_{i}^{\prime+[f]})
 {\label{eq:UDWordDiffPart1}} \\
\fat{s}_2
\Eqn{=} \sum_{i=1}^{T^f}(\fat{a}_i^{(g)}-\fat{a}_i^{\prime (g)})
+\!\!\! \sum_{i=T^f+1}^{T^f+T^g}\!\!\!\!
(\fat{d}_i^{-}-\fat{d}_i^{\prime-}){\label{eq:UDWordDiffPart2}}
\end{eqnarray}
Restricting the $f$-vector $\fat{s}_1$ of (\ref{eq:UDWordDiffPart1}) to its first $g$ components, we have
\begin{eqnarray}{\label{eq:UDWordDiffPart1_g}}
\fat{s}_1^{(g)} 
= \sum_{i=1}^{T^f}(\fat{a}_{i}^{(g)}-\fat{a}_{i}^{\prime (g)})
+\sum_{i=T^f+1}^{T^f+T^g}(\fat{d}_{i}^{+}-\fat{d}_{i}^{\prime+}).
\end{eqnarray}
Thus
\begin{eqnarray}
\fat{s}_1^{(g)} -\fat{s}_2
&=&\sum_{i=T^f+1}^{T^f+T^g}
\left(
(\fat{d}_{i}^{+}-\fat{d}_{i}^{-})
-((\fat{d}_{i}^{\prime+}-\fat{d}_{i}^{\prime-})
\right)
\nonumber \\
&=&\sum_{i=T^f+1}^{T^f+T^g}
(\fat{d}_{i}-\fat{d}_{i}^{\prime})
\end{eqnarray}
By assumption of (\ref{eq:UDWordDif0}), it holds
$$ 
\fat{s}_1=\fat{0}^f, \quad \fat{s}_2=\fat{0}^g.
$$
Then we have that
\begin{eqnarray}
\fat{0}^{g}
=\sum_{i=T^f+1}^{T^f+T^g}
(\fat{d}_{i}-\fat{d}_{i}^{\prime})
\end{eqnarray}

Since $\mathcal{D}^g$ is UD by assumption, Lemma~\ref{le:UD} implies
$$
(\fat{d}_{T^f+1}-\fat{d}_{T^f+1}^\prime, 
\dots, \fat{d}_{T^f+T^g}-\fat{d}_{T^f+T^g}^\prime,)
= \fat{0}^{gT^g}.
$$
Thus
$$(\fat{u}_{T^f+1}-\fat{u}_{T^f+1}^\prime, 
\dots, \fat{u}_{T^f+T^g}-\fat{u}_{T^f+T^g}^\prime)
=\fat{0}^{2gT^g}.
$$

From (\ref{eq:UDWordDiffPart1}) we have
$$
\sum_{i=1}^{T^f}(\fat{a}_{i}-\fat{a}_{i}^\prime)
=\fat{0}^f
$$
for which it similarly follows that
$$
(\fat{a}_{1}-\fat{a}_{1}^\prime, \dots, \fat{a}_{T^f}-\fat{a}_{T^f}^\prime) 
= \fat{0}^{fT^f}.
$$
That is 
$$(\fat{u}_{1}-\fat{u}_{1}^\prime, \dots, 
\fat{u}_{T^f}-\fat{u}_{T^f}^\prime)=\fat{0}^{2fT^f}.
$$
The proof of this lemma is completed.
\myQED

By our notation, the number of users of $\Omega(\mathcal{A}^f,\mathcal{D}^g)$ is represented by $T^{f+g}$, since its code length is $f+g$. Thus, the relationship between $T^{f}$, $T^{g}$ and $T^{f+g}$ is 
\begin{eqnarray} {\label{eq:UDTfgTfTg}}
T^{f+g}=T^{f}+T^{g}.
\end{eqnarray}
The total rate of code $\Omega(\mathcal{A}^f,\mathcal{D}^g)$ is 
\begin{eqnarray}{\label{eq:UDfgTotalRate}}
R(\Omega(\mathcal{A}^f,\mathcal{D}^g)) 
= \frac{f}{f+g}R(\mathcal{A}^f)
+\frac{g}{f+g}R(\mathcal{D}^g).
\end{eqnarray}
where 
$R(\mathcal{A}^f) = \sum\limits_{j=1}^{T^f}\frac{\log|{\mathcal{A}}_j|}{f}$
 and 
$R(\mathcal{D}^g) = \sum\limits_{j=1}^{T^g}\frac{\log|{\mathcal{D}_j}|}{g}$.

Note that when $g=0$, set $\mathcal{D}^g=\{ \emptyset \}$ is empty set. In this case, $\mathcal{A}^f=\Omega(\mathcal{A}^f,\{ \emptyset \})$.

\subsection{UD Code with Code Length $n=2^m$}
Based on the construction of (\ref{eq:UDfgCode}), we give a recursive construction of $k$-ary multiuser code $\mathcal{A}^{(n,k)}$ with code length of $n=2^m$.

For any positive integers $m$ and $k$, let
\begin{eqnarray}
k_j &=& 2^{j}(k-1)+1 \label{eq:kj} \\
\ell_j &=& \lfloor \log_{2}(k_j-1) \rfloor \nonumber \\
{\fat a}_j&=&(2^0,2^1,\cdots,2^{\ell_j-1}),\nonumber 
\end{eqnarray}
where $j=0,1, \dots, m.$ Code $\mathcal{A}^{(2^{m},k)}$ is recursively constructed as follows.

For step $i=0$, the initial codes are defined as 
\begin{eqnarray}{\label{eq:UDIniA}}
\mathcal{A}^{(2^0, k_j)}
=\{\mathcal{A}_1&=&\{ \fat{b} [{\fat a}_j^{\tt T}]\ 
|\ \fat{b} \in \{0,1\}^{\ell_j} \},
\nonumber \\
\mathcal{A}_2&=&\{0,k_j-1\}\}.\\
&& \hspace{3em} j=0,1, \dots, m. \nonumber
\end{eqnarray}

\begin{figure*}[ht]
\centering
  \includegraphics[width=18cm]{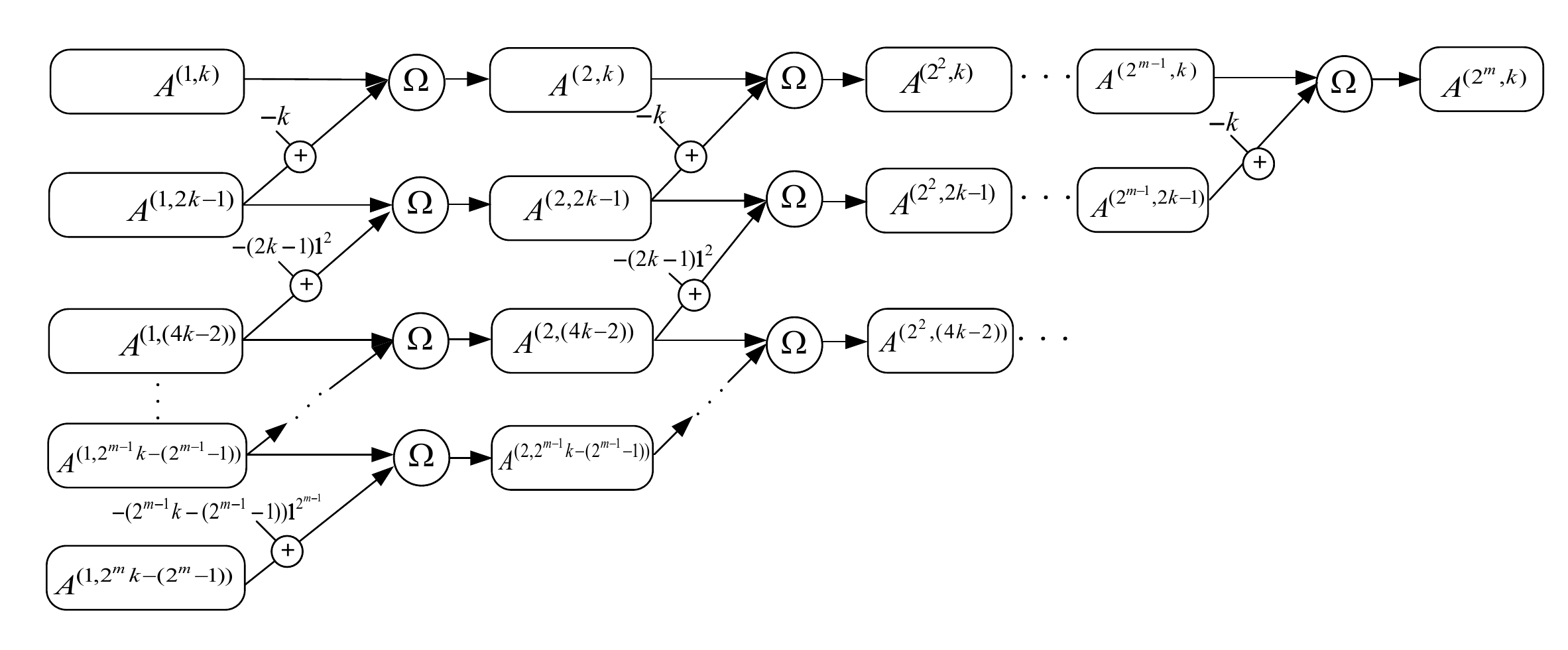}
  \vspace{-5mm}
  \caption{Recursive construction procedure of code $A^{(2^m,k)}$ from $A^{(1,k)}$, $A^{(2k-1)}$, \dots, $A^{2^{m}k-(2^m-1)}$.}   \label{figure:A2m}
\end{figure*}

Successively for steps $i=1,2,\dots, m$, using operation $\Omega(\cdot,\cdot)$ defined in (\ref{eq:UDfgCode}), we have
\begin{eqnarray} \label{eq:UDA2jAA}
\mathcal{A}^{(2^i, k_j)}
 &=& \Omega(\mathcal{A}^{(2^{i-1}, k_j)}, 
\mathcal{A}^{(2^{i-1}, k_{j+1})} - (k_{j}-1) {\bm 1}^{2^{i-1}}) \nonumber \\
&& \hspace{6em} j=0,1, \dots, m-i 
\end{eqnarray}

Therefore, from equation (\ref{eq:UDA2jAA}), we can obtain $\mathcal{A}^{(2^i, k_j)}$ at step $m$. Fig.~\ref{figure:A2m} shows the recursive construction procedure to get code $A^{(2^m,k)}$. From the initial 2-user code $\mathcal{A}^{(2^0, k_j)}$, $j=0,1,\dots,m$, by $m$ steps recursion, the code $A^{(2^m,k)}$ is obtained.

\begin{theorem}{\label{th:UD2mCode}}
Given positive integers $m$ and $k$, multi-user code $\mathcal{A}^{(2^{i=m},k_0=k)}$ is an $(2^m,\delta_a=1,\allowbreak 2^{m+1})_k$-code with
the number of users $T(\mathcal{A}^{(2^m,k)})$ and 
the total rate as 
\begin{eqnarray}
T(\mathcal{A}^{(2^m,k)}) \Eqn{=} 2^{m+1}  {\label{eq:UDTA2m}}\\
R(\mathcal{A}^{(2^{m},k)})
\Eqn{=} m/2+ (1+\lfloor \log_{2}(k-1) \rfloor). {\label{eq:UDsumrateA2m}}
\end{eqnarray}
\end{theorem}

{\em Proof}: Let $T^{(2^i,k_j)}$ be the number of users of code $\mathcal{A}^{(2^i,k_j)}$ for all $0 \leq i,j \leq m$. First, we want to show
\begin{eqnarray}{\label{eq:UDT2m}}
T^{(2^m=2^m,k_0=k)}=2^{m+1}.
\end{eqnarray}

All the initial codes of (\ref{eq:UDIniA}) have two users, i.e.,
$$
T^{(1,k_{j})}=2, \quad j=0,1,\dots,m.
$$
From  (\ref{eq:UDTfgTfTg}), we have
\begin{eqnarray*}
T^{(2^i,k_j)}
&=&T^{(2^{i-1},k_j)}+T^{(2^{i-1},k_{j+1})}\\
&=& \sum_{p=0}^{m} \binom{m}{p} T^{(2^{i-m},k_{j+p})}.
\end{eqnarray*}
When $i=m$ and $j=0$, it follows that
\begin{eqnarray*}
T^{(2^{i=m},k_0=k)}
= \sum_{p=0}^{m} \binom{m}{p} T^{(1,k_{p})}
=2^{m+1}.
\end{eqnarray*}
This verifies (\ref{eq:UDT2m}).

Next, let's show sum rate of (\ref{eq:UDsumrateA2m}). For compact notation, let $R^{(2^i,k_j)} \triequ R(\mathcal{A}^{(n^i,k_j)})$ be the sum rate of code $\mathcal{A}^{(n^i,k_j)}$. From Theorem~\ref{th:UDfgCode} and (\ref{eq:UDfgTotalRate}), we have
$$
R^{(2^i,k_j)}
=\frac{1}{2}(R^{(2^{i-1},k_j)}+R^{(2^{i-1},k_{j+1})}).
$$
By recursion, we have
\begin{eqnarray*}
R^{(2^i,k_j)}
&=&\frac{1}{2^m} \sum_{p=0}^{m} \binom{m}{p} R^{(2^{i-m},k_{j+p})}.
\end{eqnarray*}
When $i=m$ and $j=0$, it follows that
\begin{eqnarray*}
R^{(2^{i=m},k_0=k)}
&=& \frac{1}{2^m} \sum_{p=0}^{m} \binom{m}{p} R^{(1,k_{p})}
\end{eqnarray*}
where
\begin{eqnarray*}
R^{(1,k_{p})}
\Eqn{=}1+\ell_p=1+\lfloor \log_{2}(2^{p}(k-1)) \rfloor\\
\Eqn{=}p+1+\lfloor \log_{2}(k-1) \rfloor.
\end{eqnarray*}
Finally, we have
\begin{eqnarray*}
\lefteqn{R^{(n_m=2^m,k_0=k)}}\\
&=&\frac{1}{2^m}\sum_{i=0}^{m} \binom{m}{i} 
(i+1+\lfloor \log_{2}(k-1) \rfloor)\\
&=&\frac{1}{2^m}\sum_{i=0}^{m} i\binom{m}{i}
+(1+\lfloor \log_{2}(k-1)\rfloor)
\frac{1}{2^m}\sum_{i=0}^{m} \binom{m}{i}
\\
&=&m/2+(1+\lfloor \log_{2}((k-1)).
\end{eqnarray*}
Therefore we verify (\ref{eq:UDsumrateA2m}).

Finally, we show code $\mathcal{A}^{(n_m=2^m,k_0=k)}$ is UD by recursion. The initial 2-user code $\mathcal{A}^{(2^0, k_j)}$, $j=0,1,\dots,m$, of (\ref{eq:UDIniA}) is UD. Assume that $\mathcal{A}^{(2^{i-1}, k_{j})}$, $j=1,2,\dots,m-i$, is UD. Thus set 
$
\mathcal{D}^{(2^{i-1}, \pm k_j)} 
\triequ \mathcal{A}^{(2^{i-1}, k_{j+1})} - (k_{j}-1) {\bm 1}^{2^{i-1}}
$
 (see (\ref{eq:UDA2jAA})) is UD. By Theorem~\ref{th:UDfgCode}, code $\mathcal{A}^{(2^i, k_j)}$ is UD. This confirms  $\mathcal{A}^{(2^{i=m},k_0=k)}$ is UD, and completes the proof.
\myQED

\begin{remark}
\redunderline{Note that equations of (\ref{eq:UDTA2m}) and (\ref{eq:UDsumrateA2m}) are the number of users and the total code of $k$-ary UD codes with $k>3$.
When $k=2$, viewed as a specific case $\mathcal{D} = \left[
    \begin{array}{c}
       2\mathcal{A} \\
       I
    \end{array}
  \right] $.
The number of users and total rate become
\begin{eqnarray*}
T(\mathcal{A}^{(2^m,2)}) \Eqn{=} 2^{m+1}-1 \\
R(\mathcal{A}^{(2^{m},2)})
\Eqn{=} m/2+1. 
\end{eqnarray*}
They are as same as that of the code of Construction 2 of \cite{Hughes96}, Hughes and Cooper give a construction of binary($k =2$) multi-user code.}
\myQED
\end{remark}

\begin{remark}
\redunderline{The construction of $\mathcal{C}^{f+g}$ (\ref{eq:UDfgCode}) have a similar architecture as that the construction of  in \cite{Lu-ITW18}. The equation of $\mathcal{C}_{i+T_f}$  in \cite{Lu-ITW18} can be viewed as a specific case of $\mathcal{D}$ that $\mathcal{D} = \left[
    \begin{array}{c}
       2\mathcal{A} \\
       I
    \end{array}
  \right] $. 
  \myQED
}
\end{remark}

\begin{example}
In this example, we give the recursive construction of $A^{(2^{m=2},3)}$.

For $i=0$, the codes $A^{(1,3)}, A^{(1,5)}, A^{(1,9)}$ are used to be initialized. When $k_0=3$, it follows $\ell_0 = \lfloor \log_{2}(3-1) \rfloor = 1$ and ${\fat a}_0=(2^0)$. Using equation (\ref{eq:UDIniA}), we have the ternary code
$$\mathcal{A}^{(1,3)}  =  \{ \mathcal{A}^{(1,3)}_1=\{0,1\}, \mathcal{A}^{(1,3)}_2 =\{0,2\} \}.$$

Similarly, we have the $5$-ary code 
$$\mathcal{A}^{(1,5)}  =  \{ \mathcal{A}^{(1,5)}_1=\{0,1,2,3\}, \mathcal{A}^{(1,5)}_2 =\{0,4\} \}.$$
and the $9$-ary code 
$$\mathcal{A}^{(1,9)}  =  \{ \mathcal{A}^{(1,9)}_1=\{0,1,2,3,4,5,6,7\}, \mathcal{A}^{(1,9)}_2 =\{0,8\} \}.$$



Successively, at step $j=1$, we construct code
\begin{eqnarray*}
\mathcal{A}^{(2,3)} =  \Omega(\mathcal{A}^{(1, 3)}, 
\mathcal{A}^{(1, 5)}-(3-1) {\bm 1}^{2^{0}})  
\end{eqnarray*}
with constituent codes
\begin{eqnarray*}
\mathcal{A}^{(2,3)}_1  \Eqn{=} \{00,11\}, \\ 
\mathcal{A}^{(2,3)}_2  \Eqn{=} \{00,22\}, \\
\mathcal{A}^{(2,3)}_3  \Eqn{=} \{02,01,00,10\}, \\
\mathcal{A}^{(2,3)}_4  \Eqn{=} \{02,20\} 
\end{eqnarray*}


and
\begin{eqnarray*}
\mathcal{A}^{(2,5)} =  \Omega(\mathcal{A}^{(1, 5)}, 
\mathcal{A}^{(1, \pm 9)} - (5-1)\bm{1}^{2^{0}})
\end{eqnarray*}
with constituent codes
\begin{eqnarray*}
\mathcal{A}^{(2,5)}_1  \Eqn{=} \{00,11,22,33\}, \\ 
\mathcal{A}^{(2,5)}_2  \Eqn{=} \{00,44\}, \\
\mathcal{A}^{(2,5)}_3  \Eqn{=} \{04,03,02,01,00,10,20,30\}, \\
\mathcal{A}^{(2,5)}_4  \Eqn{=} \{04,40\} 
\end{eqnarray*}

At step $j=2$, 
we obtain the code $$\mathcal{A}^{(4,3)} =  \Omega(\mathcal{A}^{(2, 3)}, 
\mathcal{A}^{(2, 5)} - (3-1){\bm 1}^{2^{1}})$$ 
with constituent codes
\begin{eqnarray*}
\mathcal{A}^{(4,3)}_1 \Eqn{=} \{0000,1111\}, \\ 
\mathcal{A}^{(4,3)}_2 \Eqn{=} \{0000,2222\}, \\
\mathcal{A}^{(4,3)}_3 \Eqn{=} \{0202,0101,0000,1010\}, \\
\mathcal{A}^{(4,3)}_4 \Eqn{=} \{0202,2020\} \\
\mathcal{A}^{(4,3)}_5 \Eqn{=} \{0022,0011,0000,1100\}, \\ 
\mathcal{A}^{(4,3)}_6 \Eqn{=} \{0022,2200\}, \\
\mathcal{A}^{(4,3)}_7 \Eqn{=} \{0220,0120,0020,0021,\\ && ~0022,0012,0002,1002\}, \\
\mathcal{A}^{(4,3)}_8 \Eqn{=} \{0220,2002\} 
\end{eqnarray*}
The total rate of code is $R(\mathcal{A}^{(4,3)}) = 3 \text{~bits/channel use}.$
\myQED
\end{example}

\subsection{UD Code with Arbitray Length}
This section gives a $T^{n}$-user UD $k$-ary code with an arbitrary code length $n$ based on Theorem~\ref{th:UDfgCode} and Theorem~\ref{th:UD2mCode}.

To give a $T^n$-user UD $k$-ary code, some notations are needed. An arbitrary positive integer $n$ is represented as the binary form
\begin{eqnarray} {\label{eq:defn}}
n=\sum_{j=0}^{r}n_j2^j, \qquad n_j\in \{0,1\},
\end{eqnarray}
where
\begin{eqnarray} {\label{eq:defrnj}}
r=\lfloor \log_2n \rfloor.
\end{eqnarray}
Using the binary form $(n_r,n_{r-1},\cdots,n_0)$ of $n$, put
\begin{eqnarray} 
f_j &=& n_j2^j, {\mbox{ \ for \ }} 
j=0,1,2,\cdots,r, 
\label{eq:fj} \\
g_j &=& \sum_{i=0}^{j}n_i2^{i}, 
{\mbox{ \ for \ }} 
j=0,1,2,\cdots,r. 
\label{eq:gj}
\end{eqnarray}
Then,
\begin{eqnarray} {\label{eq:gjfjgj}}
 g_{j}=f_j+g_{j-1}, 
\ {\mbox{ for}} \ j=1,2,\cdots,r. 
\end{eqnarray}
Especially, when $j=r$,
\begin{eqnarray} {\label{eq:gr}}
n = g_{r} = f_r+g_{r-1}. 
\end{eqnarray}

Moreover, let 
\begin{eqnarray*}
\tilde{k}_j&=&2^{(\sum_{i=0}^{r}n_i-\sum_{i=0}^{j}n_i)} \cdot (k-1)+1 \\
&=& \left\{
\begin{array}{ll}
2^{\sum_{i=j+1}^{r}n_i}\cdot (k-1)+1, {\mbox{ if  }} j=0,1,2,\dots,r-1 \\
k, {\mbox{ if  }} j=r.
\end{array}
\right.
\end{eqnarray*}

We are ready to give a $T^{n}$-user UD code $\mathcal{A}^{(n,k)}$. For an arbitrary positive integer $n$, when codes $\mathcal{A}^{(2^j,\tilde{k}_j)}$ ($j=1,2,\cdots,r$) of (\ref{eq:UDIniA}) and (\ref{eq:UDA2jAA}) are given {\em a priori}, a $T^{n}$-user UD code $\mathcal{A}^{(n,k)}$ is constructed as follows:

For $j=0$, the code is initially defined by
\begin{eqnarray}{\label{eq:UDAg0Def}}
\mathcal{A}^{(g_0,\tilde{k}_0)}
= \left\{
	\begin{array}{ll}
		\mathcal{A}^{(2^0,\tilde{k}_0)},   & {\mbox{ if\ }} n_0=1, \\
		\{ \emptyset \},  & {\mbox{ if\ }} n_0=0,
	\end{array}
\right.
\end{eqnarray}
where $\mathcal{A}^{(1,\tilde{k}_0)}$ is given in (\ref{eq:UDIniA}).

Successively for $j=1,2,\cdots,r$, a code $\mathcal{A}^{(g_{j},\tilde{k}_j)}$ is constructed from $\mathcal{A}^{(2^j,\tilde{k}_j)}$ and $\mathcal{A}^{(g_{j-1},\tilde{k}_{j-1})}$ by recursion
\begin{eqnarray}\label{eq:UDAgRecursion}
\mathcal{A}^{(g_{j},\tilde{k}_j)}
= \left\{
  \begin{array}{l}
  {\Omega}\left(\mathcal{A}^{(2^j,\tilde{k}_j)},
    \mathcal{A}^{(g_{j-1},\tilde{k}_{j-1})}-\frac{\tilde{k}_{j-1}-1}{2}\fat{1}^{g_{j-1}}\right), \\
  \hspace*{13em} {\mbox{ if }} n_j=1 \\
  \mathcal{A}^{(g_{j-1},\tilde{k}_j)},\\
  \hspace*{13em} {\mbox{ if\ }} n_j=0.
  \end{array}
\right.
\end{eqnarray}
where $\mathcal{A}^{(2^j,\tilde{k}_j)}$ is given in (\ref{eq:UDfgCode}).

Finally, due to (\ref{eq:gr}), put
\begin{eqnarray}{\label{eq:UDAnAgr}}
\mathcal{A}^{(n,k)}=\mathcal{A}^{(g_{r},\tilde{k}_r)}.
\end{eqnarray}
In (\ref{eq:UDAgRecursion}), ${\Omega}(\cdot,\cdot)$ is defined in (\ref{eq:UDfgCode}) with $f=2^j$ and $g=g_{j-1}$. Note that $2^j \geq g_{j-1}$.

The number $T^{(n,k)}$ of users and the total rate of $\mathcal{A}^{(n,k)}$ are
\begin{eqnarray}
T^{(n,k)}&=&2n,
{\label{eq:UDTanyn}} \\
R(\mathcal{A}^{(n,k)})
&=&
\frac{1}{n}\sum_{j=0}^{r}n_jj2^{j-1} 
+\frac{1}{n}\sum_{j=0}^{r-1}n_j 2^j \sum_{i=j+1}^{r}n_i
\nonumber \\
&&\qquad +1+\lfloor \log_2 (k-1) \rfloor.
\label{eq:UDRanyn}
\end{eqnarray}
The derivations of (\ref{eq:UDTanyn}) and (\ref{eq:UDRanyn}) are shown in Appendix \ref{app:DerivationTn}.

\begin{theorem}{\label{th:UDTnUser}}
The $T^{(n,k)}$-user code $\mathcal{A}^{(n,k)}$ of (\ref{eq:UDAnAgr}) is {\em UD}. 
\myQED
\end{theorem}
\noindent {\em Proof }: 
The theorem is proved by induction. 

Let us observe $\mathcal{A}^{(g_0,\tilde{k}_0)}$ of (\ref{eq:UDAg0Def}). When $n_0=0$, $\mathcal{A}^{(g_0,\tilde{k}_0)}$ is the empty set. When $n_0=1$, $\mathcal{A}^{(1,\tilde{k}_0)}$ is UD.

Prove that if $\mathcal{A}^{(g_{j-1},\tilde{k}_{j-1})}$ is UD, then $\mathcal{A}^{(g_{j},\tilde{k}_{j})}$ is UD. According to (\ref{eq:UDAgRecursion}), two cases must be considered:

{\em Case 1:} $n_j=1$, it holds 
$$
\mathcal{A}^{(g_{j},\tilde{k}_{j})}
={\Omega}(\mathcal{A}^{(2^j,\tilde{k}_{j})},
\mathcal{D}^{(g_{j-1},\pm \tilde{k}_{j-1})})
$$
where 
$
\mathcal{D}^{(g_{j-1},\pm\tilde{k}_{j-1})} 
\triequ \mathcal{A}^{(g_{j-1},\tilde{k}_{j-1})}
-(\tilde{k}_{j-1}-1)\fat{1}^{g_{j-1}}
$ (see (\ref{eq:UDAgRecursion})).
Since $\mathcal{A}^{(g_{j-1},\tilde{k}_{j-1})}$ is UD over $\{0,1,\dots, (k_{j-1}-1)\}^{g_{j-1}}$,
$
\mathcal{D}^{(g_{j-1},\pm\tilde{k}_{j-1})} 
$ is UD over $\{0,\pm 1,\dots, \pm (k_{j-1}-1)\}^{g_{j-1}}$. It follows from Theorem~\ref{th:UDfgCode} that $\mathcal{A}^{(g_{j},\tilde{k}_{j})}$ is UD, since $\mathcal{A}^{(2^j,\tilde{k}_{j})}$ is UD.

{\em Case 2:} $n_j=0$,  it holds $\mathcal{A}^{(g_{j},\tilde{k}_{j})}=\mathcal{A}^{(g_{j-1},\tilde{k}_{j-1})}$. It is obvious that $\mathcal{A}^{(g_{j},\tilde{k}_{j})}$ is UD.

Therefore, $\mathcal{A}^{(n=g_r,\tilde{k}_r=k)}$ is UD. This completes the proof of this theorem.
\myQED
The derivations of (\ref{eq:UDTanyn}) and (\ref{eq:UDRanyn}) are shown in Appendix \ref{app:DerivationTn}.

\begin{remark}
\redunderline{Table \ref{tab:Trates} gives a numerical comparison of code parameters between codes of (\ref{eq:UDAnAgr}) and \cite{Lu-ITW18}. Although both codes have arbitrary code length, we compare their total rates for a fixed number of users.
The proposed codes have
the higher code rate and shorter code length than those of previous codes \cite{Lu-ITW18} for a fixed number of users.}

\begin{table}[htb]
\begin{center}
\caption{Comparison of total rates of UD $k$-ary codes with $\ell = \lfloor \log_2(k-1) \rfloor$}\label{tab:Trates}
\begin{tabular}{c||c|c|c|c} \hline \hline
numbers & \multicolumn{2}{|c|}{${\mathcal C}^n$: code in Corollary} & \multicolumn{2}{|c}{
${\mathcal C}^{n \prime}$: the code in \cite{Lu-ITW18}
} \\ \cline{2-5}
 of & code &total rates &code &total rates \\
 users &lengths & (b/c) &lengths & (b/cu) \\ \hline \hline
8 & 4 &2.000+$\ell$& 3  & 1.666+$\ell$  \\
14& 7 &2.286+$\ell$& 5  & 2.000+$\ell$  \\
20& 10&2.500+$\ell$& 7  & 2.286+$\ell$  \\
26& 13&2.692+$\ell$ & 9& 2.444+$\ell$   \\
32& 16&3.000+$\ell$& 11 & 2.545+$\ell$  \\
 \hline \hline
\end{tabular}

\end{center}
\end{table}
\end{remark}

\begin{example}
In this example, we give the recursive construction of code $A^{(7,3)}$ with code length $n=7$.

The binary form of $n=7$ is $n_2 =1, n_1 = 1, n_0 =1$. 

For $j=0$, $\tilde{k}_0 = k_{2} =9$, the code is initially defined by 
$\mathcal{A}^{(g_0,\tilde{k}_0)}
= \mathcal{A}^{(2^0, 9)}$ as shown in Example~1.

For $j=1$, $\tilde{k}_1 = k_{1} =5$,
\begin{eqnarray*}
\mathcal{A}^{(g_1,\tilde{k}_1)}
\Eqn{=} \mathcal{A}^{(3,5)} \\
\Eqn{=} {\Omega}\left(\mathcal{A}^{(2^{1},5)},
    \mathcal{A}^{(2^0,9)}-(5-1)\fat{1}^{2^0}\right) 
\end{eqnarray*}
with constituent codes as
\begin{eqnarray*}
\mathcal{A}^{(3,5)}_1  \Eqn{=} \{000,111,222,333\}, \\ 
\mathcal{A}^{(3,5)}_2  \Eqn{=} \{000,444\}, \\
\mathcal{A}^{(3,5)}_3  \Eqn{=} \{040,030,020,010,000,101,202,303\}, \\
\mathcal{A}^{(3,5)}_4  \Eqn{=} \{040,404\}, \\
\mathcal{A}^{(3,5)}_5  \Eqn{=}
\{004,003,002,001,000,100,200,300\}, \\
\mathcal{A}^{(3,5)}_6  \Eqn{=}
\{004, 400\}.
\end{eqnarray*}

For $j=2$,  $\tilde{k}_2 = k_{0} = 3$,
\begin{eqnarray*}
\mathcal{A}^{(g_2,\tilde{k}_2)}
\Eqn{=} \mathcal{A}^{(7,3)} \\
\Eqn{=} {\Omega}\left(\mathcal{A}^{(2^{2},3)},
    \mathcal{A}^{(3,5)}-(3-1)\fat{1}^{3}\right) 
\end{eqnarray*}
with constituent codes as
\begin{eqnarray*}
\mathcal{A}^{(7,3)}_1 \Eqn{=} \{0000000,1111111\}, \\ 
\mathcal{A}^{(7,3)}_2 \Eqn{=} \{0000000,2222222\}, \\
\mathcal{A}^{(7,3)}_3 \Eqn{=} \{0202020,0101010,0000000,1010101\}, \\
\mathcal{A}^{(7,3)}_4 \Eqn{=} \{0202020,2020202\} \\
\mathcal{A}^{(7,3)}_5 \Eqn{=} \{0022002,0011001,0000000,1100110\}, \\ 
\mathcal{A}^{(7,3)}_6 \Eqn{=} \{0022002,2200220\}, \\
\mathcal{A}^{(7,3)}_7 \Eqn{=} \{0220022,0120012,0020002,0120012,\\ && ~0022002,0012001,0020002,1002100\}, \\
\mathcal{A}^{(7,3)}_8 \Eqn{=} \{0220022,2002200\} \\
\mathcal{A}^{(7,3)}_9  \Eqn{=} \{0000333,0000222,0000111,0000000\}, \\ 
\mathcal{A}^{(7,3)}_{10}  \Eqn{=} \{0000333,1110000\}, \\
\mathcal{A}^{(7,3)}_{11}  \Eqn{=} \{ 0100303,0000303,0000313,0000323,\\
&& 0000333,0000232,0000131,0000030\}, \\
\mathcal{A}^{(7,3)}_{12}  \Eqn{=} \{ 0100303, 1010030\}, \\
\mathcal{A}^{(7,3)}_{13}  \Eqn{=}
\{0010330,0000330,0000331,0000332,\\
&& 0000333,0000233,0000133,0000033\}, \\
\mathcal{A}^{(7,3)}_{14}  \Eqn{=}
\{0010330, 1000033\}.
\end{eqnarray*}

The code rate is $R(\mathcal{A}^{(7,3)}) = 3.286$ bits/channel use.
\myQED
\end{example}

\section{Conclusions}{\label{sec:conclusions}}
We proposed a construction of a UD $k$-ary $T$-user coding scheme for MAAC.
We first give a construction of $k$-ary $T^{f+g}$-user UD code from a $k$-ary $T^{f}$-user UD code and a $k^{\pm}$-ary $T^{g}$-user difference set with its two component sets $\mathcal{D}^{+}$ and $\mathcal{D}^{-}$ {\em a priori}. Based on the $k^{\pm}$-ary $T^{g}$-user difference set constructed from a $(2k-1)$-ary UD code, we recursively construct a UD $k$-ary $T$-user codes with code length of $2^m$ from initial multi-user codes of $k$-ary, $2(k-1)+1$-ary, \dots, $(2^m(k-1)+1)$-ary. Introducing multi-user codes with higer-ary makes the total rate of generated code $\mathcal{A}$ higher than that of conventional code.

Using the UD $k$-ary code proposed in this paper, applying the construction in \cite{Lu-ISITA14}, error-correcting multi-user codes with higher total rate can also be constructed.

\section*{Acknowledgement}
This work was supported in part by JSPS KAKENHI Grant Number16K06373, 18K04132 and 18H0113325, in part by MEXT through the Strategic Research Foundation at Private Universities under Grant S1411030, and in part by the Fundamental Research Funds for the Central Universities of China (XJS17111).

\appendix[Derivations of $T^{n}$ in (\ref{eq:UDTanyn}) and $R(\mathcal{A}^{n})$ in (\ref{eq:UDRanyn})] {\label{app:DerivationTn}}

We first show (\ref{eq:UDTanyn}). The number $T_n$ of users in $\mathcal{A}^n$ is counted according to the recursive procedure of (\ref{eq:UDAgRecursion}). From (\ref{eq:UDTfgTfTg}) and (\ref{eq:gjfjgj}), for $j=1,2,\cdots,r$, it follows
\begin{eqnarray}{\label{eq:Tgj}}
T^{(g_{j},\tilde{k}_j)}
=n_jT^{(2^j,\tilde{k}_j)}+T^{(g_{j-1},\tilde{k}_{j-1})}. 
\end{eqnarray}
Moreover, from (\ref{eq:Tgj}) and (\ref{eq:gjfjgj}), we have
\begin{eqnarray} 
T^{(g_r=n,\tilde{k}_r)} \Eqn{=} n_rT^{(2^r,\tilde{k}_r)}+T^{(g_{r-1},\tilde{k}_{r-1})}  \nonumber \\
\Eqn{=} n_rT^{(2^r,\tilde{k}_r)}+n_{r-1}T^{(2^{r-1},\tilde{k}_{r-1})}+T^{(g_{r-2},\tilde{k}_{r-2})} \nonumber \\
    & \vdots & \nonumber \\
\Eqn{=} \sum_{j=1}^{r}n_jT^{(2^j,\tilde{k}_j)} +T^{(g_{0},\tilde{k}_0)}  \nonumber \\
\Eqn{=} \sum_{j=1}^{r}n_jT^{(2^j,\tilde{k}_j)} +n_0T^{(2^{0},\tilde{k}_0)} 
{\label{eq:UDTanyninduce}}
\end{eqnarray} 
Substituting (\ref{eq:UDT2m})into (\ref{eq:UDTanyninduce}), we obtain
\begin{eqnarray*}
T^n=\sum_{j=0}^{r}n_j2^{j+1} =2n
\end{eqnarray*}
which verifies (\ref{eq:UDTanyn}).

Second, we verify (\ref{eq:UDRanyn}). From (\ref{eq:gj}) and (\ref{eq:UDfgTotalRate}), it follows that
\begin{eqnarray}{\label{eq:UDRategj}}
&& R(\mathcal{A}^{(g_j,k_j)}) \nonumber \\
\Eqn{=}\frac{1}{f_j+g_{j-1}} \left(n_jf_jR(\mathcal{A}^{(f_j,\tilde{k}_j)})
+g_{j-1}R(\mathcal{A}^{(g_{j-1},\tilde{k}_{j-1}}) \right) \nonumber \\
\Eqn{=}\frac{1}{g_{j}}\left(n_j2^jR(\mathcal{A}^{(f_j,\tilde{k}_j)})+g_{j-1}R(\mathcal{A}^{(g_{j-1},\tilde{k}_{j-1})})\right) \nonumber
\end{eqnarray}

Moreover, from (\ref{eq:gr}) we have
\begin{eqnarray*}
&& R(\mathcal{A}^{(g_r=n,\tilde{k}_r)}) \\
\Eqn{=}\frac{1}{n}\left(n_r2^rR(\mathcal{A}^{(f_r,\tilde{k}_r)})
+g_{r-1}R(\mathcal{A}^{(g_{r-1},\tilde{k}_{r-1}})
\right) \nonumber \\
\Eqn{=}\frac{1}{n}\left(n_r2^rR(\mathcal{A}^{(f_r,\tilde{k}_r)})
+n_{r-1}2^{r-1}R(\mathcal{A}^{(f_{r-1},\tilde{k}_{r-1})}) \right. \\
&& \left. +g_{r-2}R(\mathcal{A}^{(g_{r-2},\tilde{k}_{r-2}}) 
\right) \nonumber \\
& \vdots & \nonumber \\
\Eqn{=}\frac{1}{n}\left(
\sum_{j=0}^{r}n_j2^jR(\mathcal{A}^{(2^j,\tilde{k}_j)})\right) 
 \nonumber \\
\Eqn{=}\frac{1}{n}\left(
\sum_{j=0}^{r}n_j2^j[\frac{j}{2}+(1+\tilde{\ell}_j)]
\right) \nonumber \\
\Eqn{=}\frac{1}{n}
\sum_{j=0}^{r}n_jj2^{j-1} +\frac{1}{n}\sum_{j=0}^{r}n_j2^j(1+\tilde{\ell}_j) \nonumber \\
\Eqn{=}\frac{1}{n}
\sum_{j=0}^{r}n_jj2^{j-1} +\frac{1}{n}\sum_{j=0}^{r}n_j2^j
+\frac{1}{n}\sum_{j=0}^{r}n_j2^j\tilde{\ell}_j \nonumber
\end{eqnarray*}

Since $\tilde{\ell}_j=\lfloor \log_2 (\tilde{k}_j-1) \rfloor$, it follows that
\begin{eqnarray*}
&&\frac{1}{n}\sum_{j=0}^{r}n_j2^j\tilde{\ell}_j \\
\Eqn{=} \frac{1}{n}\sum_{j=0}^{r-1}n_j 2^j (\sum_{i=j+1}^{r}n_i +  \lfloor \log_2 (k-1) \rfloor)
+\frac{n_r 2^r}{n} \lfloor \log_2 (k-1) \rfloor \nonumber \\
\Eqn{=}\frac{1}{n}\sum_{j=0}^{r-1}n_j 2^j \sum_{i=j+1}^{r}n_i+\frac{1}{n}\sum_{j=0}^{r}n_j 2^j \lfloor \log_2 (k-1) \rfloor \nonumber \\
\Eqn{=}\frac{1}{n}\sum_{j=0}^{r-1}n_j 2^j \sum_{i=j+1}^{r}n_i+\lfloor \log_2 (k-1) \rfloor.
\end{eqnarray*}
Then we verify (\ref{eq:UDRanyn}).
\myQED

\end{document}